\documentclass[a4paper,12pt]{article}
\usepackage{graphicx}
\usepackage{epsfig}
\usepackage{amsmath}
\usepackage{a4wide}

\title{Zones of quiet in a broadband diffuse sound field}
\author{Boaz Rafaely \\ Institute of Sound and Vibration Research \\ University of Southampton, Southampton, SO17 1BJ, UK}
\date{3 April, 2001}

\begin{document}

\maketitle

\paragraph{}  \begin{flushleft}
Abbreviated title: Broadband zones of quiet \\
\end{flushleft}

\paragraph{}  \begin{flushleft}
\emph{Corresponding author:}\\
Boaz Rafaely\\
Institute of Sound and Vibration Research\\
University of Southampton, \\
Southampton, SO17 1BJ\\
UK\\
Email:  br@isvr.soton.ac.uk\\
Tel:    +44 (0)23 80593043\\
Fax:    +44 (0)23 80593190\\
\end{flushleft}

\newpage
\abstract The zones of quiet in pure-tone diffuse sound fields
have been studied extensively in the past, both theoretically and
experimentally, with the well known result of the 10\,dB
attenuation extending to about a tenth of a wavelength. Recent
results on the spatial-temporal correlation of broadband diffuse
sound fields are used in this study to develop a theoretical
framework for predicting the extension of the zones of quiet in
broadband diffuse sound fields. This can be used to study the
acoustic limitations imposed on local active sound control systems
such as an active headrest when controlling broadband noise.
Spatial-temporal correlation is first revised, after which
derivations of the diffuse field zones of quiet in the near-field
and the far-field of the secondary source are presented. The
theoretical analysis is supported by simulation examples comparing
the zones of quiet for diffuse fields excited by tonal and
broadband signals. It is shown that as a first approximation the
zone of quiet of a low-pass filtered noise is comparable to that
of a pure-tone with a frequency equal to the center frequency of
the broadband noise bandwidth.

\paragraph{}
PACS numbers: 43.50.Ki, 43.55.Cs

\newpage
\section{Introduction}

\paragraph{}
Active control of sound has been studied intensively in the past
two decades, both theoretically and
experimentally$^{\ref{N92},\ref{E99}}$. Global control of sound in
enclosures was shown to be limited to the very low frequencies,
where only few acoustic modes dominate the sound
field$^{\ref{N92}}$, and so in many cases active control is
practical only locally, generating limited zones of quiet. A
typical application for local active sound control is a noise
reducing headrest in a passenger seat, attenuating noise around
the passenger's
ears$^{\ref{O53},\ref{R99JASA},\ref{R99IEEE},\ref{N00},\ref{C97},\ref{G95}}$.
Since local control would usually be performed in enclosures, a
model which was often used is that of diffuse primary sound field,
and a decaying near field to model the secondary pressure from a
closely located source. Pure-tone sound fields have been studied
extensively for such local control, developing theoretical limits
on the spatial extension of the zone of
quiet$^{\ref{E88},\ref{J94a}}$, and verifying the results with
experiments$^{\ref{G95}}$. It was shown that the 10\,dB zone of
quiet is extended to about tenth of a wavelength for pure-tone
sound fields$^{\ref{J94b}}$. The analysis of zones of quiet in
diffuse fields used the well-known spatial correlation function of
pure-tone diffuse fields$^{\ref{E88},\ref{J94a}}$, derived by Cook
\textit{et al.} in the 50's$^{\ref{C55}}$.

\paragraph{}
Although the theoretical and experimental results for pure-tone
local control were useful to predict the performance of active
headrest attenuating low frequency tonal noise in propeller
aircraft, for example, in many cases the nature of the noise is
broadband, such as in most jet passenger aircraft, and so
pure-tone results will be of limited use in this case. For a
broadband local active control system a useful measure of
performance would be the spatial extent of the overall sound
attenuation, which requires the analysis of broadband sound fields
and so cannot make use of the pure-tone results. Previous studies
of broadband local control systems were performed experimentally,
by including, for example, the effect of the feedback control
system, which would usually be used in this
case$^{\ref{R99JASA},\ref{R99IEEE}}$. It was shown that broadband
local active control could be useful in practice, although in
addition to the limitations imposed by the acoustics, other
limitations are also imposed by the control system, due to, for
example, the delay in the response between the loudspeaker and the
cancellation point.

\paragraph{}
The aim of this paper is to develop a theoretical framework for
predicting the spatial extent of the zones of quiet in broadband
diffuse sound fields. This can then be used to predict performance
limitations of broadband active headrest systems as imposed by the
acoustics, and can complement previous experimental results.
Similar to the pure-tone zones of quiet case, the analysis of
broadband zones of quiet presented here employ spatial correlation
of diffuse sound fields. However, in this work \emph{broadband}
spatial-temporal correlation is used, as developed recently by
Rafaely$^{\ref{R00}}$.

The paper is organized as follows. First the diffuse sound field
and spatial correlation are introduced, after which theoretical
results for broadband local control are developed, for near-field
control, but also for far-field control, where the secondary
source is located away from the cancellation point. Finally
simulation results for various broadband sound fields are
presented and compared to the pure-tone case. It is shown that as
a first approximation, broadband zone of quiet can be predicted
from that of tones at the mid-frequency of the broadband noise
bandwidth.

\section{The Diffuse sound field}

The plane wave model of a diffuse sound field assumes an infinite
number of plane waves, arriving uniformly from all directions,
with random phases$^{\ref{P81},\ref{J79}}$. Although a perfect
diffuse fields rarely exists, the model is widely used for
reverberant sound fields analysis, where the field is assumed to
be sufficiently diffuse. A commonly used definition for
sufficiently diffuse field is that by Schroeder$^{\ref{S96}}$,
which defines the field being diffuse above the Schroeder
frequency. This corresponds to the frequency above which there
exists at least three room modes within the 3\,dB bandwidth of any
one mode$^{\ref{P81}}$. This complements the wave model of a
diffuse field if it is assumed that each mode can be represented
by eight plane waves$^{\ref{K82}}$, and so a large number of
significant modes implies a large number of plane waves, which in
the limit approaches the definition of a perfect diffuse field.
The pressure in a perfect diffuse field can therefore be written
as a function of space and time in spherical coordinates
$\mathbf{r}=(r,\theta,\phi)$ as$^{\ref{P81}}$:

\begin{equation} \label{planewaves}
p(\mathbf{r},t) = \lim_{N \to \infty} \frac{1}{N} \sum_{n=1}^N
p_n(\mathbf{r},t)
\end{equation}

where $p(\mathbf{r},t)$ is the total pressure at position
$\mathbf{r}$ and time $t$, $N$ is the number of plane waves which
approaches infinity, and $p_n$ is the n'th plane wave. The spatial
correlation in pure tone diffuse field was studied both
theoretically and experimentally by Cook \emph{et
al.}$^{\ref{C55}}$, who showed that it behaves as a sinc function,

\begin{equation} \label{sinc}
\rho(\Delta \mathbf{r}) = \mathrm{sinc}(k \Delta \mathbf{r}) =
\frac{\sin(k \Delta \mathbf{r})}{k \Delta \mathbf{r}}
\end{equation}

where $k$ denotes the wave number, and $\rho$ the correlation
coefficient, which can be defined, assuming the sound field is
stationary over both space and time$^{\ref{N92}}$, as:

\begin{equation} \label{Ecorr}
\rho(\Delta \mathbf{r},\Delta t) =
\frac{E\left[p(\mathbf{r}_1,t_1)
p(\mathbf{r}_0,t_0)\right]}{E[p^2]}
\end{equation}

where $\Delta \mathbf{r}$ denotes the distance between the two
points, $\Delta \mathbf{r}=|\mathbf{r}_1-\mathbf{r}_0|$, $\Delta
t$ denotes the time lag given by $\Delta t = t_1-t_0$, $E[\cdot]$
denotes the expectation operation which is calculated as the
average over many samples of diffuse sound fields, and $E[p^2]$ is
the variance of the pressure which is not dependent on
$\mathbf{r}$ or $t$ due to the stationarity assumption. As
discussed above, (\ref{sinc}) was widely used in the theoretical
analysis of zones of quiet in pure-tone diffuse sound fields.
Rafaely$^{\ref{R00}}$, recently developed an expression for the
correlation which can incorporate both pure-tone and broadband
sound fields, and which depends on the power spectral density of
the signal exciting the diffuse field,

\begin{equation} \label{corr_int}
\rho(\Delta \mathbf{r},\Delta t) = \frac{1}{2\pi E[p^2]}
\int_{-\infty}^{\infty} S(\omega) \mathrm{sinc}\left(\frac{\omega
\Delta \mathbf{r}}{c}\right) e^{j\omega \Delta t} d\omega
\end{equation}

where $E[p^2]$ is equal to the integral over $S(\omega)$, i.e. the
signal power. Equation (\ref{corr_int}) enables the extension of
the pure-tone local control results to broadband sound fields, as
shown in the following sections.

\section{Near-field broadband active sound control}
\label{nearfield}

Local active sound control in a diffuse sound field can be
achieved by introducing a secondary source and cancelling the
total pressure in the near-field of the source. A simple model
used to theoretically study such an approach is that of a monopole
secondary source in a primary diffuse sound field. This
arrangement was used by Joseph \emph{et al.}$^{\ref{J94a}}$ to
study zones of quiet in pure-tone diffuse fields, and provided a
useful insight into the performance of more practical near-field
active sound control systems such as an active headrest system. A
derivation of the spatial extension of the zones of quiet in
\emph{broadband} diffuse sound field for local active control is
presented in this section. This is a novel result which can be
used to predict the spatial extent of the overall attenuation of
the broadband noise in diffuse sound fields.

\paragraph{}
Consider a secondary monopole source placed at the origin of a
spherical coordinate system, $\mathbf{r}=(r,\theta,\phi)$, with
the resulting pressure denoted by $p_s(\mathbf{r},t)$. The primary
sound field is diffuse and is denoted by $p_p(\mathbf{r},t)$. The
total pressure is a superposition of the primary and secondary
pressure contributions and is given by

\begin{equation} \label{ptotal}
p(\mathbf{r},t) = p_p(\mathbf{r},t) + p_s(\mathbf{r},t)
\end{equation}

The pressure at position $\mathbf{r}_0=(r_0,\theta_0,\phi_0)$ is
cancelled, i.e. $\mathbf{r}_0$ is assumed to be the cancellation
point, such that

\begin{equation} \label{pcancel}
p_p(\mathbf{r}_0,t) + p_s(\mathbf{r}_0,t) = 0
\end{equation}

It is now assumed that position $\mathbf{r}_0$ is in the near
field of the secondary source, such that the indirect secondary
sound field resulting from reflections is negligible. The distance
from the source at which the direct field dominates is referred to
as the "reverberation distance", which depends on the room volume
and reverberation time$^{\ref{S96}}$. The spatial extent of the
zone of quiet depends on how well the primary pressure is
attenuated around the cancellation point. The averaged squared
total pressure at position $\mathbf{r}_1=(r_1,\theta_1,\phi_1)$
near the cancellation point is therefore calculated, where the
expectation operation $E[\cdot]$ is used as a statistical average
over many samples of diffuse sound fields. The variance of the
total pressure at position $\mathbf{r}_1$ can therefore be written
using (\ref{ptotal}) as

\begin{equation} \label{Eptotal}
E[p^2(\mathbf{r}_1,t)] = E[p_p^2(\mathbf{r}_1,t)^2] +
E[p_s^2(\mathbf{r}_1,t)] + 2 E[p_p(\mathbf{r}_1,t)
p_s(\mathbf{r}_1,t)]
\end{equation}

Note that the variance of the total pressure at $\mathbf{r}_1$
depends on the variance of the primary and secondary fields at the
same point, but also on the correlation between the primary and
secondary fields at $\mathbf{r}_1$. Since we assumed in
(\ref{pcancel}) that both fields are equal with opposite phase at
the cancellation point $\mathbf{r}_0$, this correlation will
depend on how both fields change from $\mathbf{r}_0$ to
$\mathbf{r}_1$, which will be developed later. We next expand each
of the terms in equation (\ref{Eptotal}), and reformulate the
equation.

\paragraph{}
Assuming the diffuse primary sound field is stationary, such that
the variance of the pressure is the same for all $\mathbf{r}$ and
$t$, the following equality can be written

\begin{equation} \label{Epprim}
E[p_p^2(\mathbf{r}_1,t)] = E[p_p^2(\mathbf{r}_0,t)] = E[p_p^2]
\end{equation}

It is now assumed that the secondary source is generated by a
monopole point source. Although the monopole source is not an
accurate representation of more practical secondary sources such
as loudspeakers, under some assumptions the pressure produced by a
monopole behaves in a similar way to that produced by a piston in
a baffle, which is often used to model sound radiation from
loudspeakers. These assumptions are$^{\ref{B86}}$: (1) $ka<0.5$,
or $a<\frac{\lambda}{4\pi}$, which means that the source radius
$a$ is much smaller than a wavelength, and the source can
therefore be considered omni-directional, and (2) $r>a$, which
suggests that only pressure further away than one source radius is
considered. For example, these assumption will hold for a 4 inch
($a=5$\,cm) loudspeaker, for frequencies below about 500\,Hz,
further than 5\,cm from loudspeaker. These are reasonable
assumptions considering a practical local active control system
such as active headrest$^{\ref{R99JASA},\ref{R99IEEE}}$, and so
the monopole model should provide useful insight into the
behaviour of more practical systems.

\paragraph{}
The secondary sound field produced by a monopole point source in
the near-field is assumed to generate spherical waves, which
propogate away from the source and decay in
amplitude$^{\ref{K82},\ref{N90}}$:

\begin{equation} \label{psec}
p_s(r,t) = \frac{\rho_0}{4\pi r} \dot{q}\left(t-\frac{r}{c}\right)
\end{equation}

which is now dependent only on the distance from the source, $r$,
with $q$ denoting the source strength (volume velocity per unit
volume) and $\dot{q}$ its derivative with respect to time. The
secondary pressure at $\mathbf{r}_1$ can now be written in terms
of the secondary pressure at $\mathbf{r}_0$ using (\ref{psec}) as:

\begin{equation} \label{psec2}
p_s(\mathbf{r}_1,t) = \frac{r_0}{r_1}
p_s\left(\mathbf{r}_0,t-\frac{\Delta r}{c}\right)
\end{equation}

where $\Delta r = r_1-r_0$, which is the difference in the
distances of the two points $\mathbf{r}_1$ and $\mathbf{r}_0$ to
the source.

\paragraph{}
The averaged squared secondary pressure at $\mathbf{r}_1$ can now
be written using (\ref{psec2}), (\ref{pcancel}) and (\ref{Epprim})
as:

\begin{eqnarray} \label{Epsec}
E[p_s^2(\mathbf{r}_1,t)] &=& \left(\frac{r_0}{r_1}\right)^2
E\left[p_s^2\left(\mathbf{r}_0,t-\frac{\Delta r}{c}\right)\right]
\nonumber \\ &=& \left(\frac{r_0}{r_1}\right)^2
E\left[p_p^2\left(\mathbf{r}_0,t-\frac{\Delta r}{c}\right)\right]
= \left(\frac{r_0}{r_1}\right)^2 E[p_p^2]
\end{eqnarray}

The last term in (\ref{Eptotal}) can also be written using
(\ref{psec2}), (\ref{pcancel}) and (\ref{Ecorr}) as:

\begin{eqnarray} \label{Epsp}
E[p_p(\mathbf{r}_1,t) p_s(\mathbf{r}_1,t)] &=&
E\left[p_p(\mathbf{r}_1,t) \frac{r_0}{r_1}
p_s\left(\mathbf{r}_0,t-\frac{\Delta r}{c}\right)\right] \nonumber
\\ &=& -\frac{r_0}{r_1} E\left[p_p(\mathbf{r}_1,t)
p_p\left(\mathbf{r}_0,t-\frac{\Delta r}{c}\right)\right]
\nonumber
\\ &=& -\frac{r_0}{r_1} \rho\left(\Delta \mathbf{r},\frac{\Delta r}{c}\right)
E[p_p^2]
\end{eqnarray}

The variance of the total pressure at position $\mathbf{r}_1$ in
(\ref{Eptotal}) can now be written in terms of the variance of the
primary pressure by substituting equations (\ref{Epprim}),
(\ref{Epsec}) and (\ref{Epsp}) in equation (\ref{Eptotal}),

\begin{equation} \label{Eptotal2}
E[p^2(\mathbf{r}_1,t)] = E[p_p^2] + \left(\frac{r_0}{r_1}\right)^2
E[p_p^2] - 2 \frac{r_0}{r_1} \rho\left(\Delta
\mathbf{r},\frac{\Delta r}{c}\right) E[p_p^2]
\end{equation}

\begin{figure}
    \centering
    \includegraphics[width=15cm]{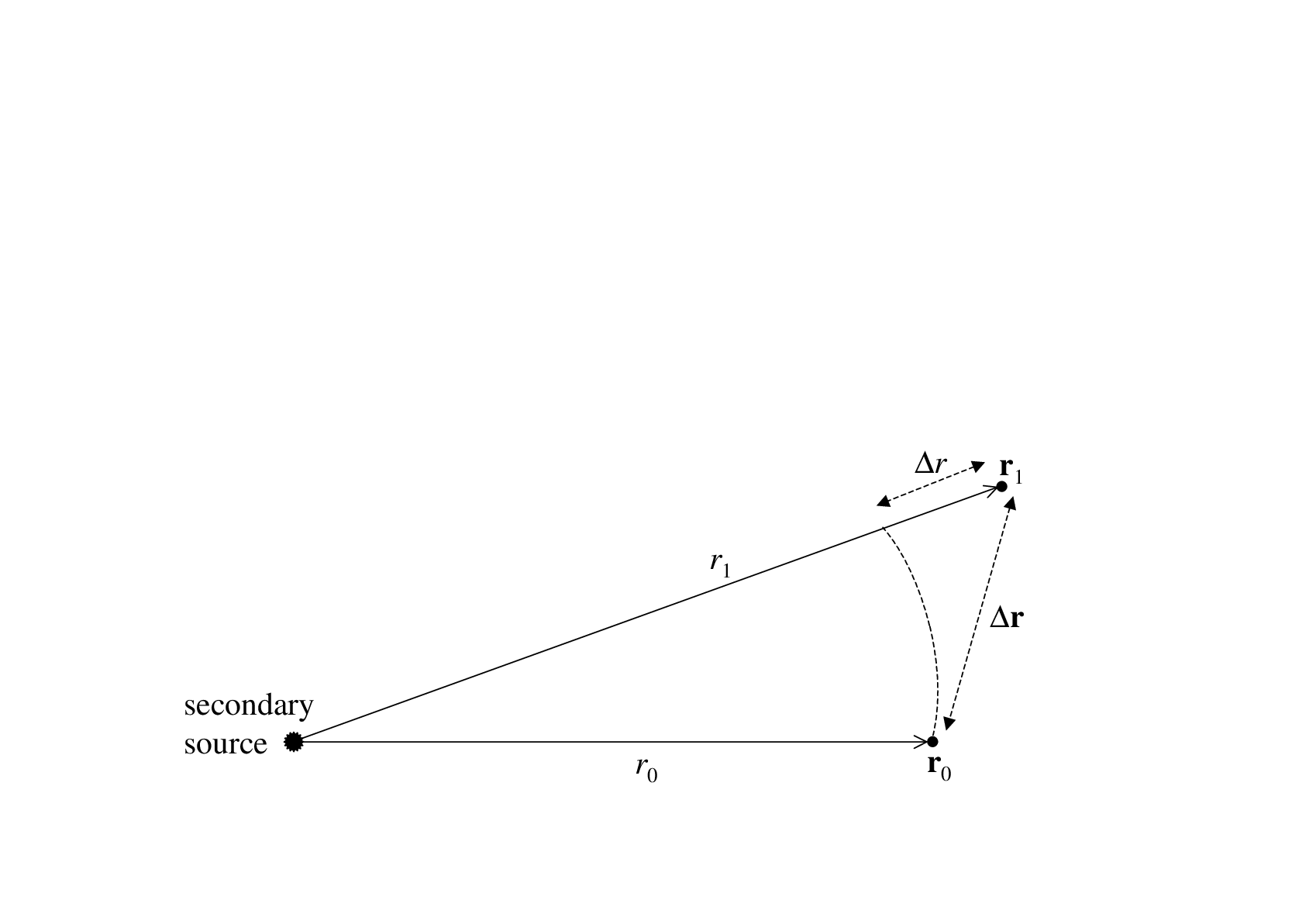}
    \caption{Graphical representation of the cancellation point, $\mathbf{r}_0$, and a position near the cancellation point, $\mathbf{r}_1$, relative to the secondary source. The distances $\Delta \mathbf{r}$ and $\Delta r$ are also illustrated in the figure.}
    \label{fig_r0r1}
\end{figure}

Dividing (\ref{Eptotal2}) by the variance of the primary pressure,
an expression for the sound attenuation $\epsilon$ at
$\mathbf{r}_1$ assuming cancellation at $\mathbf{r}_0$, is derived
as follows:

\begin{equation} \label{e}
\epsilon(\mathbf{r}_1,\mathbf{r}_0) =
\frac{E[p^2(\mathbf{r}_1,t)]}{E[p_p^2]} = 1 +
\left(\frac{r_0}{r_1}\right)^2 - 2 \frac{r_0}{r_1}
\rho\left(\Delta \mathbf{r},\frac{\Delta r}{c}\right)
\end{equation}

where the sound attenuation in dB is given by
$10\log_{10}\epsilon$. It is important to note that in (\ref{e})
$\Delta \mathbf{r} = |\mathbf{r}_1-\mathbf{r}_0|$ is the distance
from position $\mathbf{r}_1$ to the cancellation point
$\mathbf{r}_0$, while $\Delta r=r_1-r_0$ is the difference between
the distances of the two points $\mathbf{r}_1$ and $\mathbf{r}_0$
to the secondary source, as illustrated in Fig. \ref{fig_r0r1}. In
the simplified case of on-axis attenuation, the two distances are
equal, i.e. $\Delta \mathbf{r} = \Delta r$. Equation (\ref{e})
together with the expression for the spatial-temporal correlation
function in a diffuse field (equation (\ref{corr_int})), can be
used to calculate the attenuation of broadband noise in the
near-field of a monopole point source, a distance $\Delta
\mathbf{r}$ away from the cancellation point. Examples of
near-field zones of quiet in a broadband diffuse sound field are
presented below.

\section{Far-field broadband active sound control}

The previous section described active sound control in a diffuse
field where the secondary source was placed close to the
cancellation point. The latter was therefore in the near-field of
the secondary source, and the derivation that followed employed
this assumption. In this section it is assumed that the
cancellation point is far from the secondary source, such that the
resulting secondary field at the cancellation point is assumed to
be diffuse. In practice this means that the cancellation point is
further than a "revenrberation distance" or "radius of
reverberation"$^{\ref{P81}}$ away from the secondary source. In
this case both the primary and the secondary sound fields are
diffuse. Nevertheless, the two diffuse fields are assumed to be
uncorrelated, which is achieved in practice if the primary and
secondary sources are positioned sufficiently far away from each
other (more than a wavelength away for pure-tone
fields$^{\ref{N92}}$).

\paragraph{}
Unlike the case of near-field sound control which provides an
insight into the performance of practical active sound control
systems, such as an active headrest, broadband sound control using
a secondary source in the far-field is less practical. This is
because a practical feedforward control system will require a good
reference of the noise signal in advance, which is rarely
available for broadband noise, e.g. jet turbulence noise, while a
feedback control system will have poor performance due to the long
delay from the secondary source to the cancellation point. It is
important to note that such a limitation is not applicable to
pure-tone sound fields where system delay does not affect
performance. In addition, placing the secondary source far from
the cancellation point could result in large increase in the
pressure at other locations in the enclosure$^{\ref{N92}}$, which
is an undesirable side-effect. Although of less practical
relevance, the derivation of far-field broadband active sound
control is presented here for theoretical completeness.

\paragraph{}
Joseph$^{\ref{J90}}$ derived an equation for the average mean
square pressure away from the cancellation point under similar
conditions but when a pure-tone sound field was
assumed$^{\ref{N92}}$,

\begin{equation} \label{Ep_far}
E[p^2(\mathbf{r}_1)] = \left( E[p_p^2] + E[p_s^2] \right)
\left(1-\rho^2(\Delta \mathbf{r}) \right)
\end{equation}

The sound attenuation can now be derived, by dividing
(\ref{Ep_far}) with the variance of the primary pressure

\begin{equation} \label{e_far}
\epsilon(\Delta \mathbf{r}) =
\frac{E[p^2(\mathbf{r}_1)]}{E[p_p^2]} =
\left(1+\frac{E[p_s^2]}{E[p_p^2]}\right) \left(1-\rho^2(\Delta
\mathbf{r})\right)
\end{equation}

Elliott \emph{et al.}$^{\ref{E88}}$ noted that the value of
$E[p_s^2]/E[p_p^2]$ can only be defined in statistical terms, and
does not have a finite mean value. In practice, however, the
secondary source strength will be limited, and in an example
simulation$^{\ref{E88}}$, a value of $E[p_s^2]$ was used which is
three time larger than $E[p_p^2]$, and so for this example
(\ref{e_far}) can be written as:

\begin{equation} \label{e_far2}
\epsilon(\Delta \mathbf{r}) = 4 \left(1-\rho^2(\Delta
\mathbf{r})\right)
\end{equation}

Equation (\ref{corr_int}) can now be used in (\ref{e_far2}) to
compute the attenuation or the extent of the far-field zones of
quiet for a broadband sound field. Examples of such zones of quiet
are presented below.

\section{Examples of near-field zones of quiet}

\paragraph{}
Examples of near-field zones of quiet calculated using the results
derived above are presented in this section. The primary field is
assumed to be diffuse while the secondary field is excited by a
monopole point source. The cancellation point where both fields
are equal but opposite in phase is located in the near field of
the monopole source. A pure-tone diffuse sound field, which has
been well studied previously, is compared to broadband diffuse
sound fields using the results derived in this work. The diffuse
sound fields in the examples presented here are excited by the
signals as described in Table 1.

\newpage{}

\begin{tabular}{ p{2.5cm} p{10cm} }
\hline \bf{Signal} & \bf{Description} \\ \hline 300\,Hz Tone & A
300\,Hz pure tone
\\ \hline 300\,Hz LPF & Broadband signal generated by passing white noise
through a $32^{nd}$ order Butterworth low-pass filter with a
cut-off frequency of 300\,Hz \\ \hline 600\,Hz LPF & Broadband
signal generated by passing white noise through a $32^{nd}$ order
Butterworth low-pass filter with a cut-off frequency of 600\,Hz
\\ \hline BPF & Broadband signal generated by passing white noise
through an $8^{th}$ order Butterworth low-pass filter with a
cut-off frequency of 400\,Hz, and another $2^{nd}$ order
Butterworth high-pass filter with a cut-off frequency of 600\,Hz,
as used by Rafaely \emph{et al.}$^{\ref{R99JASA}}$ to analyze the
performance of a laboratory active headrest system \\ \hline
\end{tabular}

\paragraph{}
Table 1. Description of the signals used in the simulation
examples.

\begin{figure}
    \centering
    \includegraphics[width=10cm]{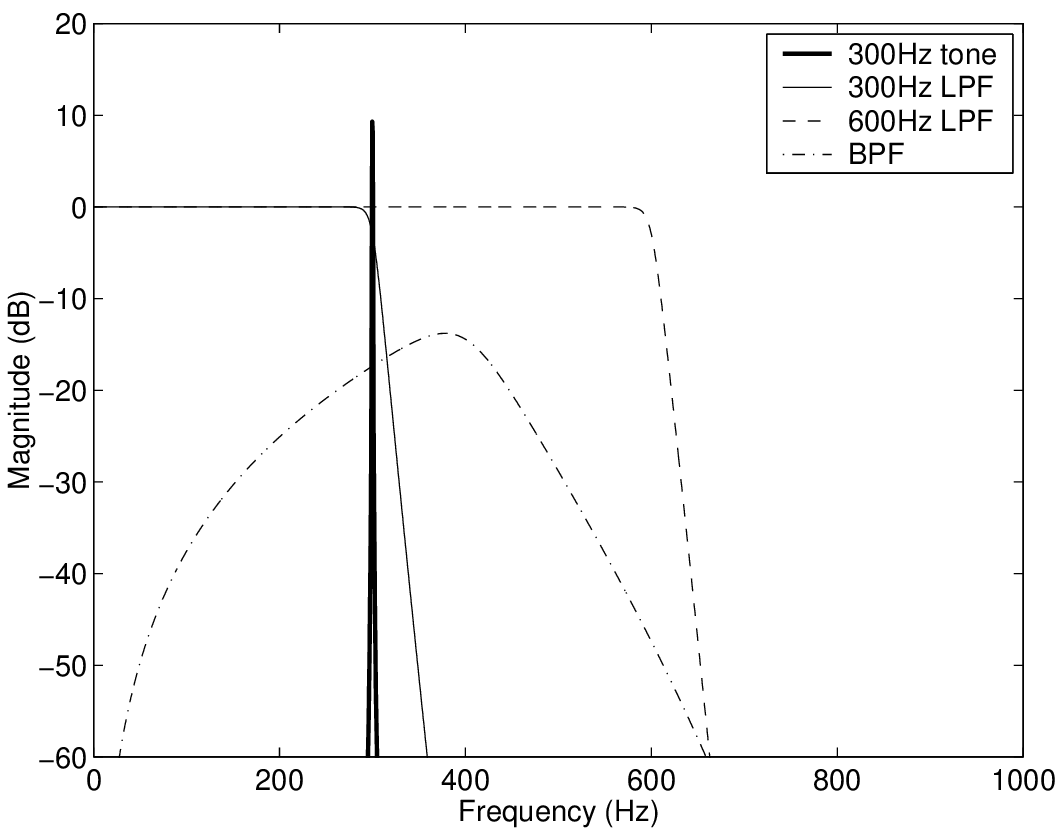}
    \caption{Power spectral density of the signals used in the simulations as described in Table 1.}
    \label{fig_spec}
\end{figure}

\paragraph{}
Figure \ref{fig_spec} shows the power spectral density of the
signals used in the simulation examples as described in Table 1.
The spatial correlation of the various primary diffuse sound
fields are compared next, after which the correlation functions
between the primary and secondary sound fields away from the
cancellation point are evaluated, which then leads to a comparison
of the zones of quiet. The spatial-temporal correlation function
for the pressure in a diffuse sound field is calculated in MATLAB
using (\ref{corr_int}) by generating the appropriate signals,
sampled at $F_s=2$\,kHz, with discrete power spectral densities
calculated using the discrete Fourier transform (DFT) having
$M=4096$ points. The integral in (\ref{corr_int}) was approximated
by a summation over frequency, as follows:

\begin{equation} \label{corr_sum}
\rho(\Delta \mathbf{r},\Delta t) \approx
\frac{1}{\sum_{m=0}^{M-1}S(m)} \sum_{m=0}^{M-1} S(m)
\mathrm{sinc}\left(\frac{2\pi m F_s}{M} \frac{\Delta
\mathbf{r}}{c}\right) e^{j \frac{2\pi m F_s}{M} \Delta t}
\end{equation}

\begin{figure}
    \centering
    \includegraphics[width=10cm]{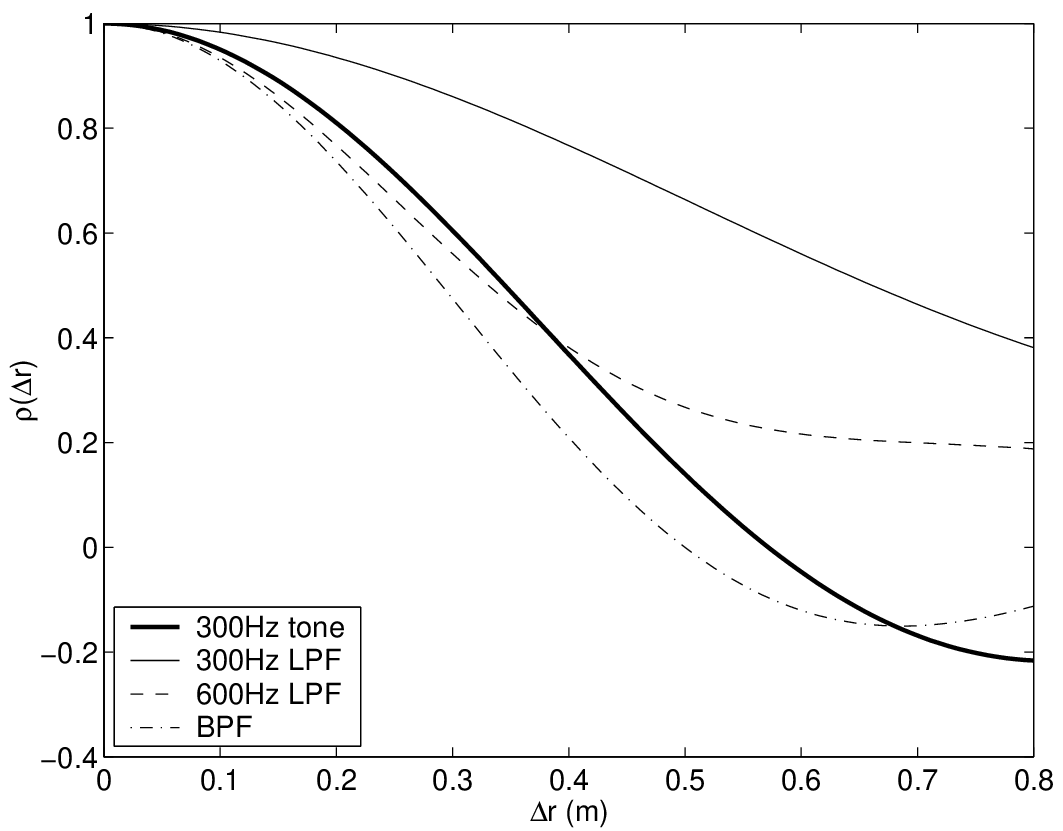}
    \caption{The spatial correlation of the primary pressure in a diffuse sound field, for the excitation signals described in Table 1.}
    \label{fig_rho_p}
\end{figure}

Figure \ref{fig_rho_p} shows the spatial correlation of the
primary diffuse field $\Delta \mathbf{r}$ away from the
cancellation point, evaluated using (\ref{corr_sum}) as $\rho
(\Delta \mathbf{r},0$), for the sound fields described in Table 1.
The figure shows that the spatial correlation for the 300\,Hz
pure-tone sound field behaves as a spatial sinc function, as
expected$^{\ref{C55}}$, with the 600\,Hz low-pass filtered noise
having similar correlation for small $\Delta \mathbf{r}$. This
observation that the spatial correlation for a band of frequencies
can be approximated by that of a pure tone at the center frequency
has been previously observed$^{\ref{C55},\ref{N97}}$. The 300\,Hz
low-pass filtered noise has higher spatial correlation, as
expected, since it is composed of lower frequencies. It is also
interesting to note that the sound field composed of the band-pass
filtered noise has a similar spatial correlation to the 600\,Hz
low-pass filtered noise, for small $\Delta \mathbf{r}$, since it
has a similar bandwidth.

\begin{figure}
    \centering
    \includegraphics[width=10cm]{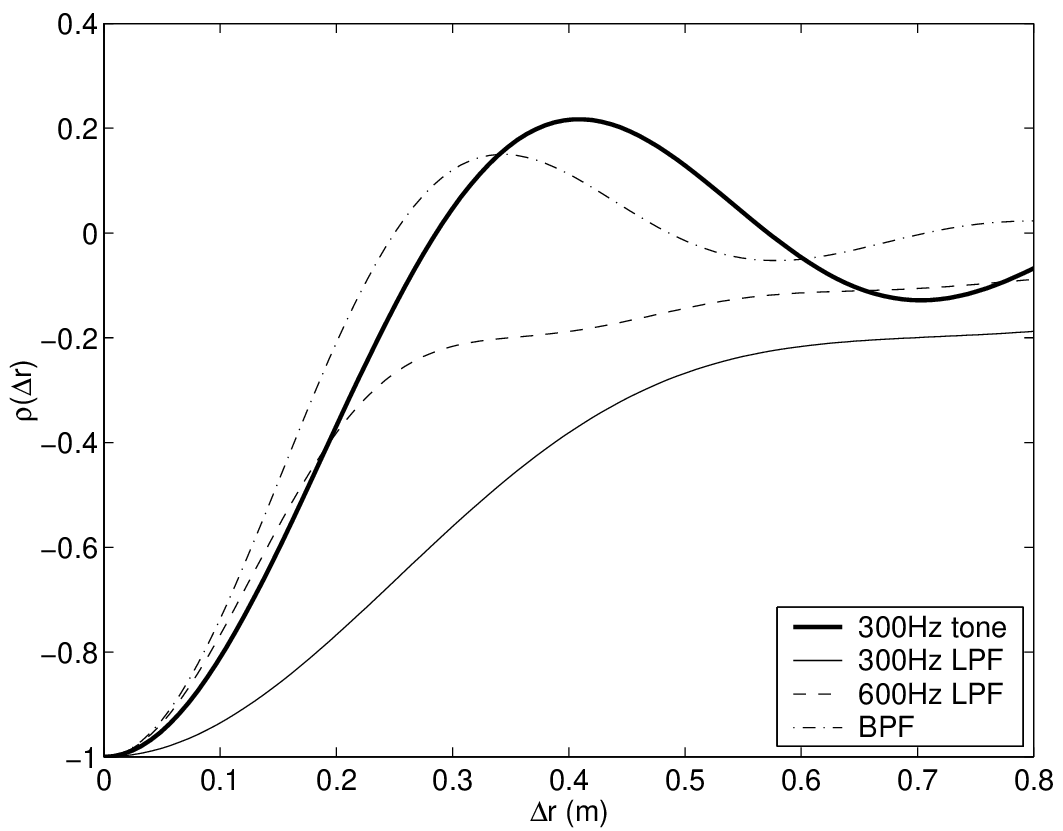}
    \caption{The spatial cross-correlation function between the primary and the secondary pressures $\Delta \mathbf{r}$ away from the cancellation point, $-\frac{r_0}{r_1} \rho(\Delta \mathbf{r}, \frac{\Delta r}{c})$, assuming $\frac{r_0}{r_1} \approx 1$ for the excitation signals described in Table 1.}
    \label{fig_rho_ps}
\end{figure}

\paragraph{}
As shown in (\ref{Eptotal}), the cross-correlation between the
primary diffuse field and the secondary near field when evaluated
at $\mathbf{r_1}$, i.e. $\Delta \mathbf{r}$ away from the
cancellation point, is used in the calculation of the total
pressure and then the attenuation at $\mathbf{r_1}$. This
cross-correlation is evaluated here for the sound fields described
in Table 1, using (\ref{Epsp}) and (\ref{corr_sum}), by
substituting $\Delta \mathbf{r}$ and $\Delta t=\frac{\Delta r}{c}$
in the spatial-temporal correlation function of the primary
diffuse field. Figure \ref{fig_rho_ps} shows this
cross-correlation for the signals described in Table 1, where it
was assumed that the cancellation point is sufficiently far from
the secondary source such that $\bf{r_1} \approx \bf{r_0}$ in
order to present the limit of the correlation values. The figure
shows that the correlation values are negative for small $\Delta
\bf{r}$ since the primary and secondary fields are equal but with
opposite phase at $\bf{r_0}$. Also, comparing the results to Fig.
\ref{fig_rho_p}, it is clear that the cross-correlation between
the primary and the secondary sound fields at $\Delta \bf{r}$ away
from the cancellation point is smaller than the auto-correlation
of the primary diffuse field for a spacing of $\Delta \bf{r}$.
This can be explained by the fact that when moving from position
$\mathbf{r}_0$, where both fields are equal with opposite phase,
to position $\mathbf{r}_1$, the primary field reduces correlation
according to (\ref{corr_int}), while the secondary field also
reduces correlation according to the near-field behaviour
described in (\ref{psec2}). The total equivalent spacing between
the two fields is therefore $2\Delta \bf{r}$, compared to only
$\Delta \bf{r}$ in Fig. \ref{fig_rho_p}, resulting in a greater
reduction in the cross-correlation compared to diffuse field
auto-correlation.

\begin{figure}
    \centering
    \includegraphics[width=10cm]{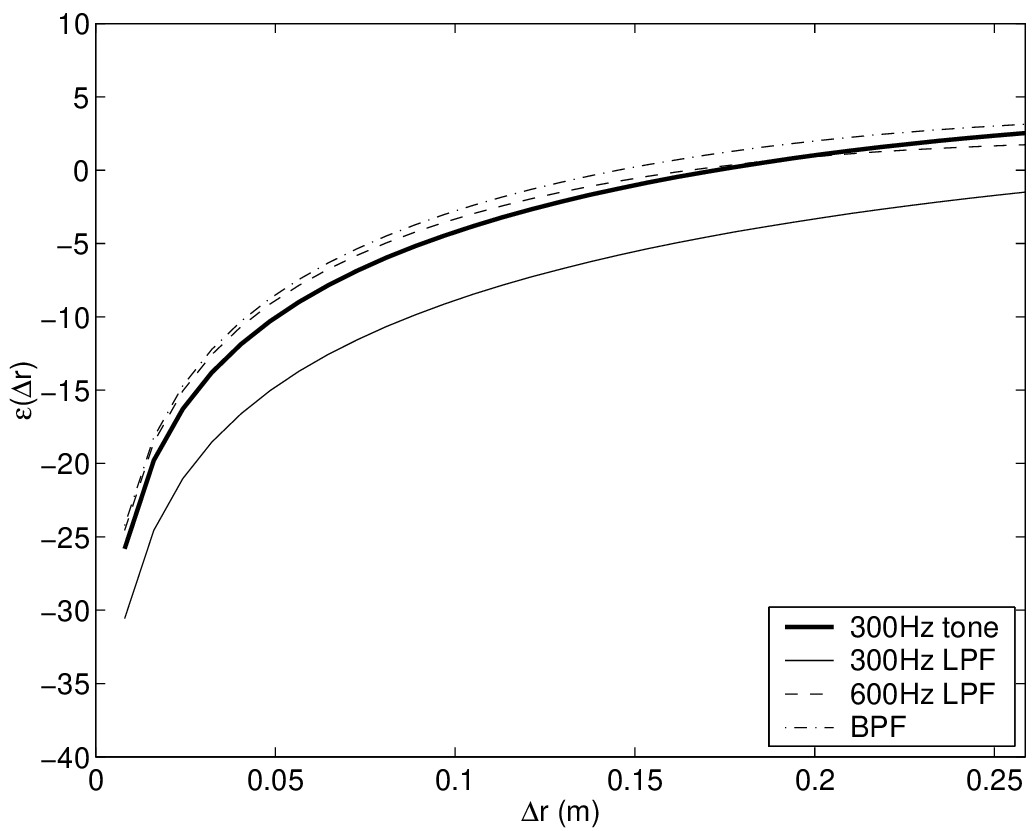}
    \caption{Attenuation as a function of the distance from the cancellation point $\Delta \bf{r}$ for the signals described in Table 1.}
    \label{fig_att_1D}
\end{figure}

\paragraph{}
The attenuation $\Delta \bf{r}$ away from the cancellation point
can be calculated using (\ref{e}) and (\ref{corr_sum}). Figure
\ref{fig_att_1D} shows the calculated attenuation for the sound
fields as described in Table 1, as a function of the distance from
the cancellation point $\Delta \bf{r}$. Again, it was assumed that
$\bf{r_1} \approx \bf{r_0}$ to present the limits of the
attenuation values. The figure shows that the 300\,Hz pure-tone
has similar zone of quiet to the 600\,Hz low-pass filtered noise
and the band-pass filtered noise, whereas the 300\,Hz low-pass
filtered noise shows larger zones of quiet. It is important to
note that the size of the zone of quiet for the 300\,Hz pure-tone,
defined by $2\Delta \bf{r}$ for $\epsilon = 0.1$, i.e. 10\,dB
attenuation, is about $0.088 \lambda$, which is slightly smaller
than the $0.1 \lambda$ rule derived by Nelson and
Elliott$^{\ref{N92}}$. This is explained by the fact that in the
derivation presented in Nelson and Elliott$^{\ref{N92}}$ the
change in the secondary sound field around the cancellation point
was approximated by a first-order function, with higher orders
neglected, whereas in this work no such approximation was made.

\begin{figure}
    \centering
    \includegraphics[width=9.5cm]{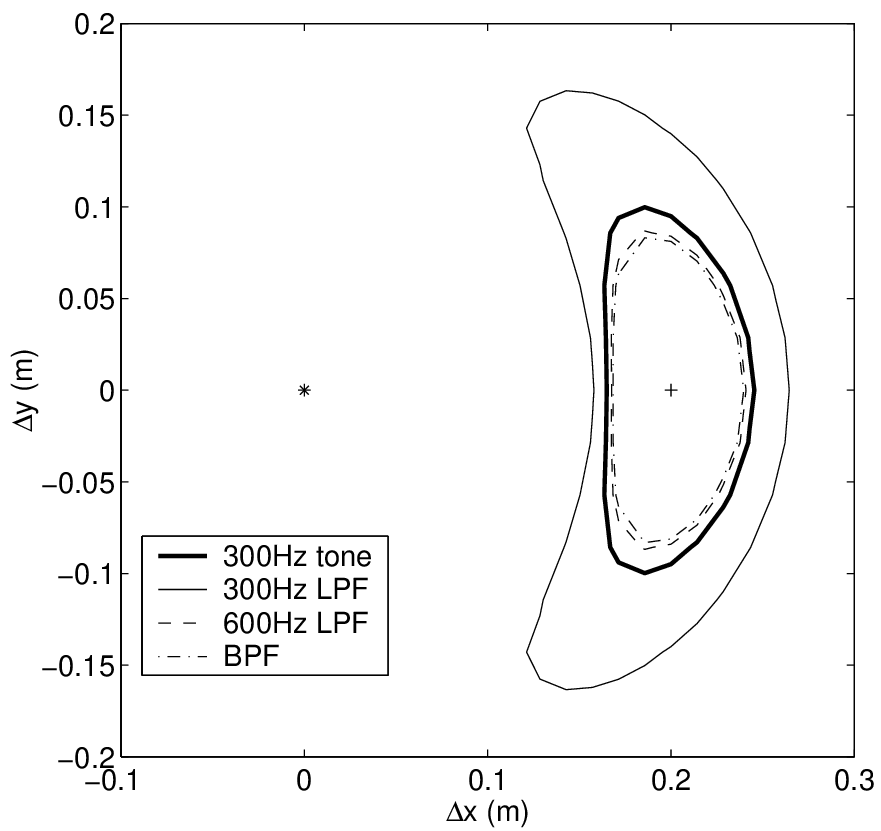}
    \caption{The 10\,dB attenuation contours as a function of $\Delta x$ and $\Delta y$, for the signals described in Table 1, with the secondary source denoted by $'*'$ and the cancellation point denoted by $'+'$ located at $(0.2,0)$.}
    \label{fig_att_2D}
\end{figure}

\paragraph{}
The 10\,dB zone of quiet is presented next, which is the
attenuation contour with a 10\,dB value. Figure \ref{fig_att_2D}
shows the calculated 10\,dB attenuation contours, or
two-dimensional zones of quiet for the sound fields described in
Table 1. In this case the monopole secondary source is located at
the origin, while the cancellation point $\bf{r_0}$ is positioned
at $(0.2,0)$, i.e. 20\,cm away from the source. The attenuation as
a function of position was calculated using (\ref{e}) and
(\ref{corr_sum}), with only the 10\,dB attenuation contour shown.
The result for the 300\,Hz tone is comparable with that of
Garcia-Bonito and Elliott$^{\ref{G95}}$, while again it is clear
that the 300\,Hz tone has similar zone of quiet to the 600\,Hz
low-pass filtered noise and the band-pass filtered noise. These
results suggest that the size of the zone of quiet for a broadband
noise of a given bandwidth, will be similar to that of a pure tone
at the middle frequency range of the broadband noise.
Nevertheless, the zone of quiet for a more general spectrum can be
calculated more accurately as described above.

\paragraph{}
The 10\,dB zone of quiet for the band-pass filtered noise is shown
to be about 8\,cm. This is slightly higher but comparable to the
zone of quiet presented by Rafaely and Elliott$^{\ref{R99JASA}}$,
for a laboratory active headrest system, which used a more
realistic experiment including a loudspeaker as a source, a
Manikin as a head, and feedback control to generate the secondary
source signal.

\section{Examples of far-field zones of quiet}

\paragraph{}
An example of far-field zones of quiet are presented in this
section, where it is assumed that the secondary source is placed
far from the cancellation point, such that both the primary field
and the secondary field are diffuse. Similar signals as for the
previous example were used here to excite the sound fields, which
are described in Table 1. Equations (\ref{e_far2}) and
(\ref{corr_sum}) with $\Delta t=0$ were used to calculate the
spatial correlation and then the attenuation for the diffuse sound
fields in this example.

\begin{figure}
    \centering
    \includegraphics[width=10cm]{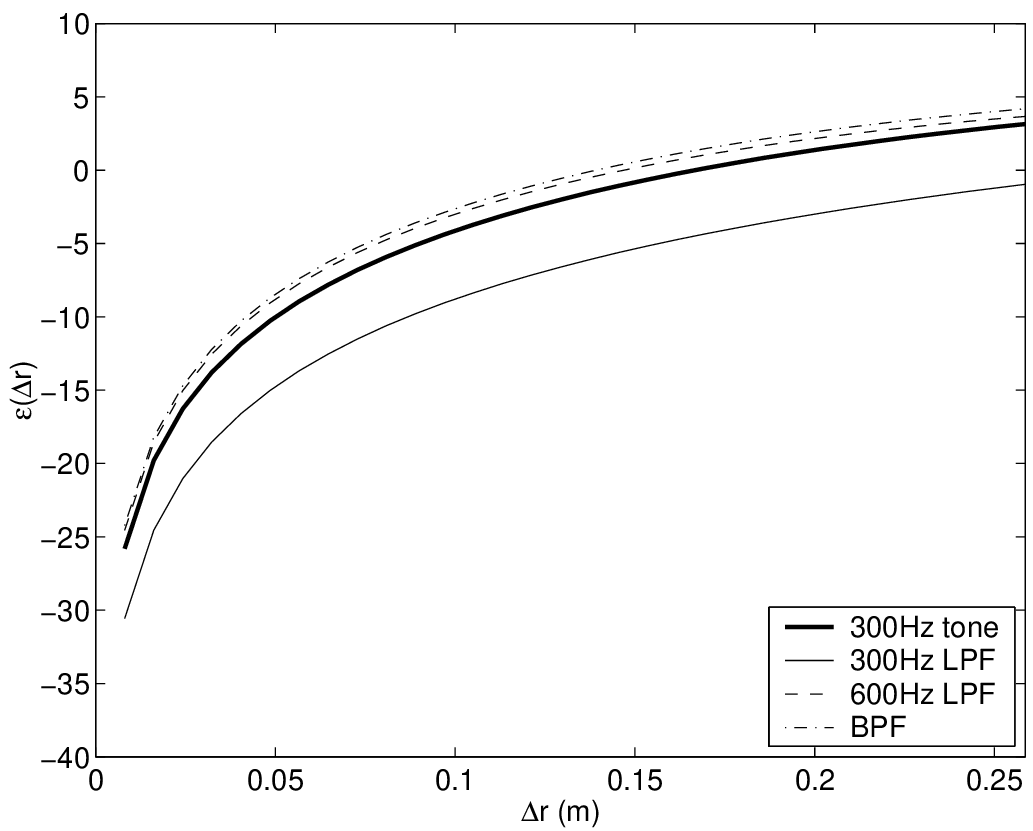}
    \caption{Far-field attenuation as a function of $\Delta \bf{r}$, for the signals described in Table 1.}
    \label{fig_att_far}
\end{figure}

\paragraph{}
Figure \ref{fig_att_far} show the attenuation as a function of
distance from the cancellation point for all four diffuse sound
fields. Results are very similar to the near-field case, and here,
as well, the zone of quiet for the broadband noise can be
approximated by that of a tone at the middle frequency.

\section{Conclusions}
The zones of quiet for broadband diffuse sound fields were derived
theoretically and then demonstrated using simulation examples.
Both near-fields zones of quiet, where the cancellation point is
in the near-field of the secondary source, and far-field zones of
quiet, where the cancellation point is in the far-field of the
secondary source, were considered. The paper demonstrated how to
calculate the zones of quiet for sound fields excited by broadband
signals, and has presented examples with several low-pass and
band-pass type random signals, comparing these to the well known
results for tonal excitations. It was shown that for simple
low-pass filtered noise, the spatial correlation and the zone of
quiet are comparable to those of a tone at the middle bandwidth
frequency. Simulation results for near-field zones of quiet for a
band-pass noise were comparable to a previous
experiment$^{\ref{R99JASA}}$ which used experimental study with a
laboratory headrest system. The theory and tools developed here
could be used to simulate and predict broadband zones of quiet
more accurately in more realistic acoustic configurations which
include real sources, a head and a control system, for example,
but this is suggested for future work.

\newpage

\end{document}